\documentclass[sigconf]{acmart}

\usepackage{floatflt}
\usepackage{wrapfig}

\AtBeginDocument{%
  \providecommand\BibTeX{{%
    \normalfont B\kern-0.5em{\scshape i\kern-0.25em b}\kern-0.8em\TeX}}}

\setcopyright{rightsretained}
\copyrightyear{2023}
\acmMonth{December}
\acmYear{2023}

\begin{document}

\title{From Prompt Engineering to Prompt Science\\ With Human in the Loop}

\author{Chirag Shah}
\orcid{0000-0002-3797-4293}
\email{chirags@uw.edu}
\affiliation{
  \institution{University of Washington, Seattle, WA, USA}
}

\renewcommand{\shortauthors}{Shah}

\begin{abstract}
As LLMs make their way into many aspects of our lives, one place that warrants increased scrutiny with LLM usage is scientific research. Using LLMs for generating or analyzing data for research purposes is gaining popularity. But when such application is marred with ad-hoc decisions and engineering solutions, we need to be concerned about how it may affect that research, its findings, or any future works based on that research. We need a more scientific approach to using LLMs in our research. While there are several active efforts to support more systematic construction of prompts, they are often focused more on achieving desirable outcomes rather than producing replicable and generalizable knowledge with sufficient transparency, objectivity, or rigor. This article presents a new methodology inspired by codebook construction through qualitative methods to address that. Using humans in the loop and a multi-phase verification processes, this methodology lays a foundation for more systematic, objective, and trustworthy way of applying LLMs for analyzing data. Specifically, we show how a set of researchers can work through a rigorous process of labeling, deliberating, and documenting to remove subjectivity and bring transparency and replicability to prompt generation process.
\end{abstract}

\ccsdesc[500]{Computing methodologies~Information extraction}
\ccsdesc[500]{Human-centered computing~User centered design}

\keywords{Large Language Models (LLMs), Prompt engineering, Qualitative coding, Verification}

\maketitle
\section{Introduction}
Large Language Models (LLMs), in the recent years, have become more sophisticated and capable for them to be applicable in many situations and tasks. These tasks are not limited to information extraction and synthesis \cite{lindemann2023sealed}, but also expanded to analysis, creation, and reasoning \cite{chew2023llm}. Unsurprisingly, many researchers have found their uses for various research tasks. LLMs are being used in identifying relevant papers \cite{paroiu2023asking}, synthesizing literature reviews \cite{antu2023using}, writing proposals \cite{gomez2023confederacy}, and analyzing data \cite{shen2023shaping}. They have been found effective and useful in investigative tasks such as drug discovery \cite{vert2023will, savage2023drug}, and while such capabilities and wide applicability of LLMs have opened up new avenues for supporting research, there is a growing concern that a large portion of this success hinges on prompt engineering, which is often an ad-hoc method to revise prompts being fed into an LLM to achieve desired response or analysis \cite{reynolds2021prompt}.

Using such a process for scientific research could be dangerous. At best, there is a possibility of creating a feedback loop with self-fulfilling prophesy. At worst, one may be overfitting hypotheses to data, leading to untrustworthy claims and findings. In most cases, over-reliance on prompt engineering could lead to unexplainable, unverifiable, and less generalizable outcomes \cite{patel2023creatively, envisioning}, lacking a scientific approach to research.

Here, we consider a scientific approach to be the one that is well-documented and transparent, verifiable, repeatable, and free of individual biases and subjectivity. We argue that most prompt engineering techniques do not lead to a scientific approach to research as they violate at least one of these principles. While such ad-hocness can be bad in most situation, when it comes to scientific research that expects a certain level of vigor, it is especially problematic.

If we want to continue harnessing the powers of LLMs, we need to do this in a responsible manner with enough scientific rigor. 
For the purpose of this work, we will consider research areas and applications where an LLM is used for labeling text data. Examples involve identifying user tasks or intents, extracting sentiments, and deriving issues from user reports.
Building on an LLM's ability to summarize or synthesize data, many recent research projects have deployed them for such tasks \cite{reynolds2021prompt, shen2023shaping}. There are also efforts to use LLMs for generating datasets and benchmarks \cite{thomas2023large, peng2023generating}. These could be foundations for future research. Therefore, it is paramount that such outcomes from LLMs are thoroughly validated and proven trustworthy.

In this article, we demonstrate how to do this by turning an ad-hoc method of prompt engineering to verifiable and replicable method of prompt science. Specifically, we take inspiration from a rich literature and methods of qualitative coding to employ human-in-the-loop for verifying (1) responses from LLMs; and (2) systematically constructing prompts.

\section{Prompt engineering prospects and problems}
LLMs are built on transformer architecture that contain encoders and decoders modules. They are trained on large amounts of text corpora. A lot of what an LLM generates heavily depends on the inner workings of this architecture and the data used for training. Most users and researchers do not have the ability to directly change that architecture or the training process. However, they can apply various techniques such as fine-tuning \cite{hu2023llm, ding2023parameter} and prompt engineering \cite{strobelt2022interactive} to shape the outputs as they desire.

\begin{figure}[htbp]
 \centering
 \includegraphics[width=0.95\linewidth]{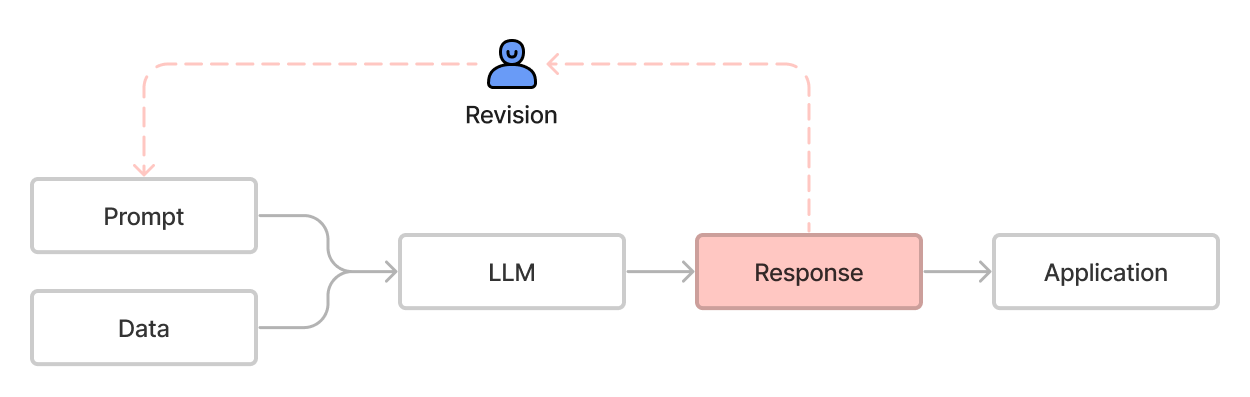}
 \caption{A typical flow of prompt engineering for research process that involves labeling data.}
 \label{fig:prompt_engineering}
\end{figure}

Figure \ref{fig:prompt_engineering} shows a conceptual flow of a typical prompt revision or engineering in a research setting. Here, a researcher is interested in getting some input data analyzed or labeled for a downstream application. To get the desired outcomes, the set of instructions or the prompt provided to the LLM is interatively revised. This has been used by many recent works such as \cite{shah2023creation, ziems2023can, chew2023llm, ma2023insightpilot}. While this could accomplish the task, it leaves us with many potential issues. For instance, if the prompt is revised by a single researcher in ad-hoc manner just to fine-tune the output, there may be personal biases and subjectivity introduced. In addition, if this process is not well-documented, it may be hard to understand or mitigate these issues. This may also make it harder for someone to verify or replicate the process. These issues may hinder scientific progress that relies on systematic knowledge being generated, questioned, and shared. There may also be issues of self-fulfilling prophecy or feedback loop that are especially dangerous when we are dealing with highly opaque systems such as an LLM.

When using LLMs for scientific research, we need some level of reliability, objectivity, and transparency in the process. Specifically, there are three problems we need to address: (1) individual bias and subjectivity; (2) vague criteria for assessment of bending the criteria to match LLM's abilities; and (3) narrow focus to a specific application instead of a general phenomenon. For this, we propose the following steps.


\begin{enumerate}
    \item Ensure that at least two qualified researchers are involved in the whole process.
    \item Before focusing on revising the prompt to produce desirable outcomes, clearly and objectively articulate what makes a desirable outcome. This will involve establishing a reliable and if possible, community-validated metric for assessment.
    \item Discuss individual differences and biases with a goal of rooting them out and creating assessments and a set of instructions that can be understood and carried out by any researcher with a reasonable set of skills to be able to replicate the experiments and achieve similar results.
\end{enumerate}

While a few recent works have shown promise in addressing these desiderata, they still lack enough rigor to meet the requirements of transparency, objectivity, and replicability. For instance, Shah et al. \cite{chen2023prompting} use a verifying method for prompt generation, but their approach lack verification of criteria as well as replicability for different datasets.

\section{Methodology}
In this section we describe a new methodology for making prompt engineering efforts to a scientific process that ensures reliability and generalizability. The previous section described how a typical prompt engineering process works; essentially, it involves revising the prompt to an LLM in order to fit the desired response and/or the downstream application. We identified that one of the problems with this endeavor is that it is quite ad-hoc, unreliable, and not generalizable. Even with enough documentation of how the final prompt was generated, one could not ensure that the same process will yield same quality output in a different situation -- be it with a different LLM or the same LLM in a different time. To overcome these problems, we describe an approach that first iterates over the criteria for assessing LLM's response and then iterates over the prompt in a systematic way using human in the loop. Our method is inspired by a typical process of qualitative coding.

\subsection{Qualitative coding with multiple assessors}
The problem we are trying to address here is that of construction of a prompt or a set of instructions that produces desired outcomes for unseen data. This is similar to building a codebook in qualitative research. Take for instance, work described in \cite{mitsui2016extracting}. Here, the authors were attempting to label user behaviors (data) with user intents (discrete labels). For this, they needed to identify a finite (and preferably a small) set of labels, and their desired descriptions. This is called a codebook.

\begin{figure}[htbp]
 \centering
 \includegraphics[width=0.95\linewidth]{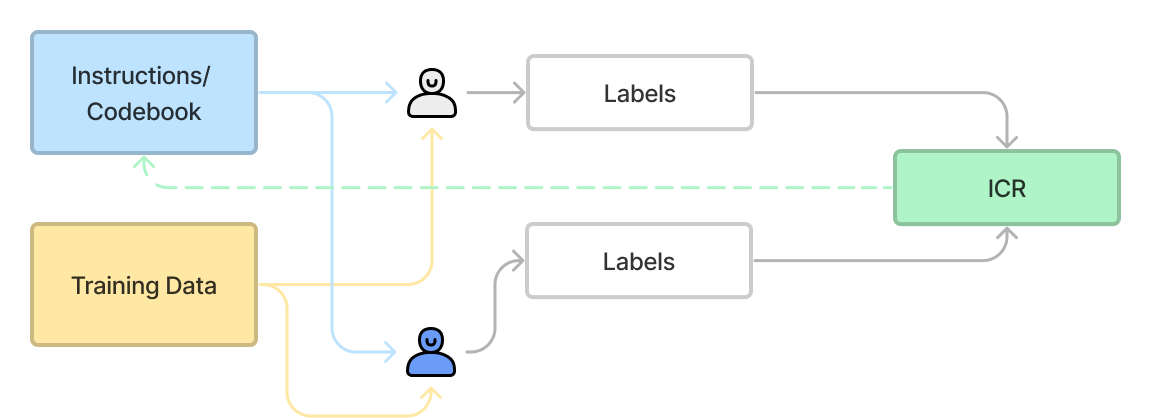}
  \captionof{figure}{Training portion of qualitative coding.}
 \label{fig:coding_training}
\end{figure}

\begin{figure}[htbp]
 \centering
 \includegraphics[width=0.95\linewidth]{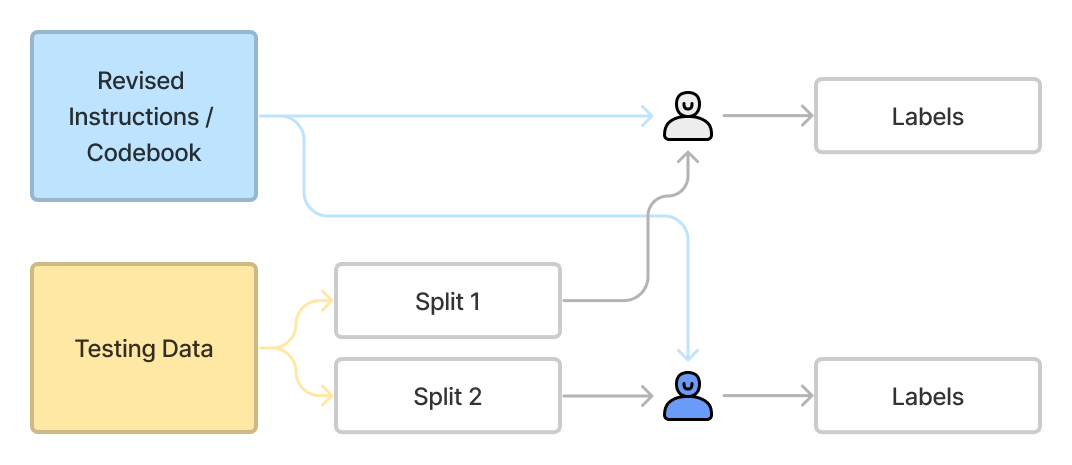}
  \captionof{figure}{Testing portion of qualitative coding.}
 \label{fig:coding_testing}
\end{figure}

They began their process by having an initial codebook that was inspired by prior work and literature. But now they need to fine-tune and validate that codebook for the given application. For this, they followed the qualitative coding process depicted in Figures \ref{fig:coding_training} and \ref{fig:coding_testing}.

As shown, for a typical qualitative coding process two researchers, who are sufficiently familiar with the task, use the initial codebook and a small set of data to independently provide labels during the training portion. The process continues until there is good enough agreement or inter-coder reliability (ICR). Once that point is reached, the codebook is considered to be ready for use. It can now be used by the same or other researchers to independently and non-overlapping label unseen data during the testing portion.

It is important to note that the value of this iterative process with human-in-the-loop. During this process, it is not just the codebook that is getting revised, but also the understanding in those two researchers. Through their discussions of disagreements, they are able to think through the task, the data, and the best ways to describe that data. This enhanced understanding impacts the codebook and vice versa. In other words, this process is used not only to produce, but also to learn and discover for the researchers. But this is not a simple form of brainstorming. The researchers are discussing their differences and charting their way forward bound by a rigorous set of criteria. For instance, in this example, the researchers were aiming to produce a codebook with categories that were clear, concise, comprehensive, and reliable. They had specific definitions of these criteria, which they used during their deliberations.

In short, such a process of qualitative coding ensures that individual subjectivity is dissolved to the extend possible while producing a coding scheme that is robust, verifiable, and explainable. Next, we show how this can be used for constructing a method for prompt generation and verification.

\subsection{Codebook building as a way to construct prompts}
\label{sec:method}
\begin{figure*}[htbp]
 \centering
 \includegraphics[width=0.95\linewidth]{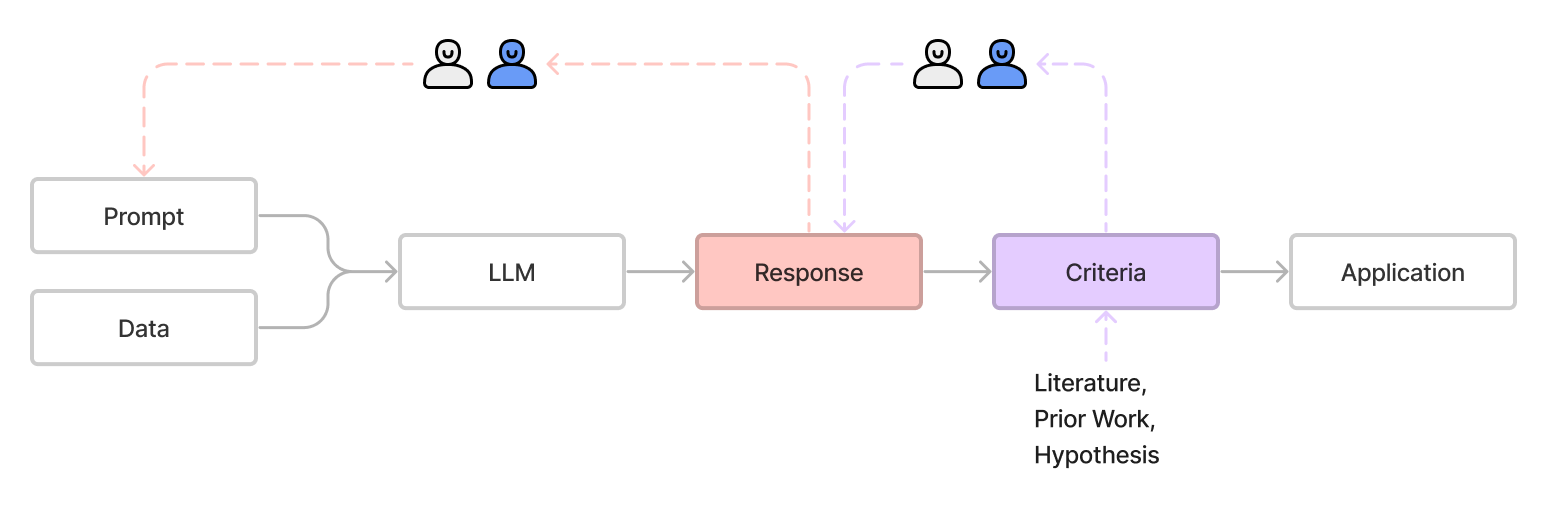}
 \caption{Proposed methodology with multi-phase approach to response and prompt validations.}
 \label{fig:pipeline}
\end{figure*}

We will now describe a new method for prompt generation inspired by the qualitative codebook building process described above. Once again, our objective is to generate a prompt for a given LLM in order to achieve desired response. In addition, we need this prompt to be systematically created with enough generalizability and explainability to be useful for same or similar experiments by other researchers, by other LLMs, and in different times. We will describe the abstract process in this section and its operationalization in the next section. To begin, we list two primary desiderata:
\begin{enumerate}
    \item The desired response from the LLM must be verified along a set of dimensions and for large enough sample size by at least two humans.
    \item The prompt to generate must consistently generate the desired responses and must be justifiable, explainable, and verifiable.
\end{enumerate}
To operationalize these two criteria, we propose a multi-phase process, working backward in the pipeline of prompt to application as shown in Figure \ref{fig:pipeline}. There are four phases to go through in this method, each explained below.

\noindent
{\em Phase 1: Set up the initial pipeline.} This can be done using anything that is reasonable. For instance, the initial prompt could be a simple and direct instruction to generate the desired response.

\noindent
{\em Phase 2: Identify appropriate criteria to evaluate the responses.} Before we can tweak the prompt, it is important to establish that we know how to evaluate the response. 
Establishing the criteria for assessing the outputs outside of prompt tuning process ensures that the researchers are focused on the task and not the LLM's capabilities.
A set of appropriate criteria could be obtained from prior work or literature. But these criteria still need to be tested and validated for the given application. Moreover, eventually we will need human assessors to assess if the LLM is generating desired responses in order to revise the prompt. For these reasons, we need to execute our first human-in-the-loop assessment. To do so, follow the steps below.

\begin{enumerate}
    \item Run the existing pipeline to generate a reasonable number of responses. Give each assessors these responses, the existing set of criteria (codebook), and sufficient details of the application. The assessors should review these things before starting their assessment. This can be supervised by the lead researcher or an expert. Note that due to the interactive learning and collaboration that takes place among the human assessors as described in Section 3.1, they do not need to have special qualifications other than being able to understand the task at hand and discuss their disagreements.
    \item Each assessor applies the criteria on the given responses independently.
    \item Compare the assessors' outcomes and compute the level of agreement. This could be using as simple as measuring the \% times when they agree, or using inter-coder reliability (ICR) using an appropriate method such as Cohen's kappa \cite{warrens2015five} or Krippendorfff's alpha \cite{krippendorff2011computing}.
    \item If the amount of agreement is not sufficient (this could vary from application to application), bring the assessors together to discuss their disagreement. This is a very important step as it offers an opportunity to not only remove individual biases and subjectivity, but also influences researchers' collective understanding of the task and the criteria for evaluation. The discussion could result in resolution of conflict, but also changes in the codebook. These changes could be about the number or the description of the criteria for assessment. Now, a new set of responses are generated for assessment using the revised codebook and the above process is repeated.
    \item Once a sufficient level of agreement is achieved, the codebook for assessing the responses is finalized. At this point the assessors may still choose to make minor changes in the codebook to enhance its readability with the idea that anyone who did not participate in this process of developing it can still understand and use it very much how those assessors would.
\end{enumerate}

\noindent
{\em Phase 3: Iteratively develop the prompt.} Continuing to work backward, now we come to the next phase of codebook development, which happens to be the first phase of the pipeline depicted in Figure \ref{fig:pipeline}. The objective here is to see if the LLM, with the current version of the prompt, is meeting our criteria for the responses. If not, we need to figure out what changes can be made to the prompt.

At this point, we know what makes a good response from our LLM. More importantly, we have now trained at least two researchers in reasonably objectively, consistently, and independently evaluating such responses. They could more effectively assess what instructions could help produce desired responses. Follow the steps below to iteratively develop the prompt.

\begin{enumerate}
    \item Run a reasonable number of data inputs using the current prompt.
    \item Give the generated responses to the human assessors, ideally the same ones who were involved in the previous phase. However, this is not a requirement. These assessors should use the codebook produced by the previous phase to assess the responses.
    \item Have them assess if the generated responses meet the criteria for the application in question.
    \item Compare the assessments by different assessors to find their level of agreement. If that agreement is not sufficient, have the assessors discuss their disagreements, possibly under the supervision of a senior researcher or an expert, and revise the codebook as necessary. Here, the codebook is the prompt fed to the LLM. Repeat the above process.
    \item Once there is enough agreement among the assessors that the prompt meets the objective of generating desired responses more than some threshold, this process is done. The assessors or the supervising researcher may still make minor adjustments to the prompt for improved readability, interpretability, and generalization.
\end{enumerate}

\noindent
{\em Phase 4: Validate the whole pipeline.} As an optional and final validation phase, run through the whole pipeline using a portion of the test data and evaluate random samples of the responses to ensure the whole process still yields quality results that can be independently and objectively validated. For this, it is ideal to use a different set of assessors and have them independently label that same set of randomly sampled responses. Compute their ICR on this sample to see if (1) there is a good enough agreement among them on labeling; and (2) the generated responses are meeting the desired criteria. If either of these objectives are not met, take appropriate actions for correction, based on phases 2 and 3.

As one can see, there are multiple checkpoints and validations in this method that ensures removal of ad-hocness, subjectivity, and opaqueness as much as possible. A proper documentation of its execution will yield a more informative, meaningful, and scientific communication that can help other researchers to trust the outputs as well as test and build such pipelines themselves.

\section{Experiments}
As a demonstration of the methodology explained in the previous section, we will now walk through its execution for two specific research applications.

\subsection{Identifying user intents}
One of the common problems in the areas of information retrieval and recommender systems is that of understanding user intents \cite{zhang2023efficiently}. Doing so requires having a taxonomy of user intents and then using such a taxonomy to label user queries, questions, or behaviors with appropriate intents. However, as shown by Shah et al. \cite{shah2023taking}, it is very challenging to construct and use such intent taxonomies. Taking one of the existing taxonomies could help with the former, but may not be sufficiently comprehensive or accurate for the latter. Similarly, creating a customized taxonomy for a given application could help applicability, but would incur a high cost.

LLMs could help in constructing and applying taxonomies, but are faced with the same challenges described in Section 1 -- namely, a lack of reliability and validation, a danger of creating feedback loops, and issues with replicability and generalizability. To overcome these issues, the author, with the help of a set of collaborators, applied the method proposed here. The following are the specific steps executed based on the pipeline presented in Figure \ref{fig:pipeline} and described in Section 3.2.

\noindent
{\em Phase 1: Set up the initial pipeline}. A pipeline that included an initial prompt and a small sample of log data from Bing Chat that could create a user intent taxonomy (zero-shot approach) was set up for initial testing.

\noindent
{\em Phase 2: Identify appropriate criteria to evaluate the responses.} We identified five criteria -- comprehensiveness, consistency, clarity, accuracy, and conciseness -- for evaluating the quality of a user intent taxonomy using prior work \cite{raad2015survey} as well as human assessments as shown in Figure \ref{fig:pipeline}.

\noindent
{\em Phase 3: Iteratively develop the prompt.} We first ran the zero-shot approach to obtain initial taxonomy. Two researchers familiar with the task of creating and using a user intent taxonomy assessed the generated taxonomy through those five criteria. Appropriate revisions were made in the original prompt and a new sample of data was passed through the LLM (here, GPT4) to create a new version of the taxonomy. The researchers repeated the process until a good agreement (inter-coder reliability or ICR \cite{cohen1960coefficient}) was obtained. 

\noindent
{\em Phase 4: Validate the whole pipeline.} Once the taxonomy was finalized (see details in \cite{shah2023using}), we proceeded with the testing phase of the pipeline. In this phase, the LLM was given new data and asked to use the human-validated taxonomy to label user intents. A set of two different researchers independently assigned their own labels using the same taxonomy. We then compared the labels among the human annotators as well as humans and LLM. We found high enough ICR for all of them. Furthermore, we repeated this process with two other LLMs -- Mistral and Hermes -- both open-sourced, and found their labeling also had strong agreements with those obtained from GPT4 as well as from human annotators. This indicates that both the taxonomy and the process for using that taxonomy through an LLM were robust and stable.

At this point, we have not only a reliable and robust pipeline for generating and using a user intent taxonomy, but it is also validated through a rigorous process with human-in-the-loop, making this trustworthy and generalizable. The details of this work, along with results and their discussions, can be found in \cite{shah2023using}.

\subsection{Auditing an LLM}
We now turn to a different type of problem involving LLMs, specifically that of auditing an LLM. While there have been many recent efforts in this area \cite{mokander2023auditing, rastogi2023supporting}, we still lack systematic and scientifically rigorous approaches. We attempted to address this using a simple, but effective method -- probing the same question differently to the given LLM to see how differently it responds. Different versions of the same question could be generated by humans, but that would hinder the scalability. Using an LLM (ideally, a different LLM than the one being audited) could help as shown in \cite{rastogi2023supporting}, but like before, we are faced with the challenge of ensuring that the generation of those probes is reliable, robust, and generalizable. To meet these challenges, we ran through the proposed method in Section 3.2.

\noindent
{\em Phase 1: Set up the initial pipeline}. We started with setting up the initial version of the pipeline. The initial prompt was set as ``Generate five different questions for the following question that represent the same meaning, but are different." for the LLM, we used Mistral and the data for original questions came from TruthfulQA dataset \cite{lin2021truthfulqa}. Given the application -- generating different versions of the same question for auditing an LLM -- we set the criteria for responses generated by Mistral as relevance and diversity.

\noindent
{\em Phase 2: Identify appropriate criteria to evaluate the responses.} We used two researchers with a good understanding of the application. Using the existing literature and the task at hand, it was determined that a good set of probes will have two important characteristics: (1) relevance or semantic similarity of each probe with the original question; and (2) good enough diversity among the probes. 
As the two researchers started independently assessing the generated probes from a small set of data (10 original questions that resulted in 50 probes) and then comparing their labels with each other, two key things happened: they started firming up the definitions of `relevance' and `diversity', and they started getting more agreements with their assessments. We had to go through three rounds of this exercise to finalize our definitions of the two criteria as well as how to label them.

\noindent
{\em Phase 3: Iteratively develop the prompt.} We wanted to have a prompt that consistently generates outputs (a set of questions) that could achieve high enough relevance and diversity for their use in a downstream application. We set the threshold for this to be 75\%. The same two researchers marked the generated responses for relevance and diversity using a new set of input data. The initial round of assessment showed that only around 50\% of the outcomes met the goal of relevance (medium or high) and diversity. To change this, the original prompt was modified to include more details about the application and the criteria. Then this process was repeated. After two more iterations, the LLM was generating outcomes that were relevant and diverse for at least 80\% of the cases.

\noindent
{\em Phase 4: Validate the whole pipeline.} A senior researcher tested the whole pipeline with the final versions of the prompt and the criteria with a few rounds of random samplings. They made a few minor adjustments to the prompt for improved readability and generalizability.
The details of these experiments, along with a demo, can be found in \cite{amirizaniani2024developing} and \cite{amirizaniani2024auditllm}.

\section{Conclusion}
LLMs could simply be stochastic string generation tools \cite{bender2021dangers} that are effective at predicting next tokens. But for many researchers, they are proving to be useful in many research problems that in general involve labeling or classifying input data based on some analysis or reasoning. However, using LLMs blindly for such research tasks just because their outputs seem reasonable can be dangerous. It can perpetual biases \cite{shah2022situating}, create fabricated or hallucinated responses \cite{rawte2023survey}, and provide unverifiable results \cite{stechly2023gpt}. All of these are harmful for scientific progress.

Creatively or iteratively designing prompts to produce desirabile outcomes, what is often referred to as prompt engineering, does not address these concerns. In fact, prompt engineering may even worsen the problems by focusing too much on bending the process to generating the desired outcomes than to developing a verifiable, validated, and replicable process. In this article, we showed how prompt engineering can be turned into prompt science by introducing a multi-phase process with human-in-the-loop. Specifically, we divided the prompt to application pipeline into two major phases that provide two levels of verification: one for the criteria for evaluating the responses generated by the LLM, and the other for the prompt for generating the responses. Through a method inspired by qualitative coding for codebook development, we showed how to have scientific rigor in developing a reliable prompt and getting trustworthy response for a downstream application. This method has been demonstrated in two different applications -- generating and using a user intent taxonomy and auditing an LLM.

We believe there are three main reasons this method provides scientific rigor.

\begin{enumerate}
    \item It provides a systematic and scientifically verifiable way of producing prompt templates, while removing individual subjectivity and biases.
    \item Through the involvement of multiple researchers documenting their deliberations and decisions, it fosters openness and replicability. This documentation can and should be made available, much like the code for an open source tool, allowing other researchers to validate, replicate, and extend the work.
    \item In addition to producing a prompt template that can generate desired outcomes reliably and consistently, the process also adds to our knowledge about the problem at hand (application), how to best assess what we need for that problem, and how we could consistently and objectively analyze given data.
\end{enumerate}

Such a rigorous process comes at a cost. Comparing the typical process of prompt engineering (Figure \ref{fig:prompt_engineering}) to the proposed method of prompt science (Figure \ref{fig:pipeline}), there is at least a 3-fold cost increase. Guided deliberations and detailed documentation can also add to the cost. Finally, the optional phase 4 calls for a lightweight, but more continual cost to the whole pipeline to maintain its quality and validity. We can compare this with an assembly line that is primarily run through automated processes with some human supervision. Depending on the risk of bad quality and the cost for manual quality check, one could decide how often and how rigorously humans should be involved. This then allows us to create more automated systems with enough control and quality assurance -- going from {\em copilot} to {\em autopilot} mode with LLMs. The key, as argued and demonstrated here, is to do so with enough human agency and scientific rigor.

Even if the goal is to not turn into a fully automated pipeline, considering that the proposed method not only leads to a outputs with higher quality, consistency, and reliability, but also an increased understanding of the task and the data, we believe the added cost is justifiable. The broader scientific community could also benefit from exercising and perhaps insisting on such a rigor so we could leverage LLMs ethically and responsibly in our research.

\begin{acks}
The taxonomy generation and usage experiments described in Section 4.1 are results of collaboration with Ryen White and other researchers from Microsoft as listed in \cite{shah2023using}. The LLM auditing experiments described in Section 4.2 are results of hard work by Maryam Amirizaniani, Adrian Lavergne, Elizabeth Okada, and Jihan Yao. We are also thankful to Aman Chadha and Tanya Roosta for their valuable comments and guidance during this process.
\end{acks}

\newpage

\bibliographystyle{ACM-Reference-Format}
\bibliography{manuscript}

\end{document}